\documentclass[twocolumn,aps,prl,amsmath,amssymb,superscriptaddress]{revtex4-2}

\usepackage{newtxtext}
\usepackage{newtxmath}
\usepackage{microtype}
\usepackage{wasysym}
\usepackage{graphicx} 
\usepackage{dcolumn}
\usepackage{bm}
\usepackage{multirow}
\usepackage{amssymb}
\usepackage{graphicx}
\usepackage{xcolor}
\definecolor{TighnariBrown}{RGB}{141,87,41}
\definecolor{TighnariGreen}{RGB}{40,114,70}
\definecolor{TighnariYellow}{RGB}{247,191,99}
\definecolor{PRLBlue}{RGB}{46,48,146}
\usepackage[
    colorlinks,
    citecolor=PRLBlue,
    linkcolor=PRLBlue,
    urlcolor=PRLBlue,
]{hyperref}
\usepackage{indentfirst}
\usepackage{cases}

\begin{document}
\title{Initial-State-Dictated Dynamical Universality Classes in Quantum Criticality}
\author{Yu-Rong Shu}
\email{yrshu@gzhu.edu.cn}
\affiliation{School of Physics and Materials Science, Guangzhou University, Guangzhou 510006, China}
\author{Shuai Yin}
\email{yinsh6@mail.sysu.edu.cn}
\affiliation{Guangdong Provincial Key Laboratory of Magnetoelectric Physics and Devices, Sun Yat-sen University, Guangzhou 510275, China}
\affiliation{School of Physics, Sun Yat-sen University, Guangzhou 510275, China}

\date{\today}

\begin{abstract}
  The concept of universality class lies at the core of phase transition physics, with each class uniquely characterized by a dedicated set of critical exponents. Here, we introduce the concept of initial-state-dictated dynamical universality classes in nonequilibrium quantum critical dynamics. By studying the imaginary-time relaxation dynamics of the dimerized Heisenberg model, we demonstrate that distinct kinds of initial states can give rise to different dynamical universality classes within the same equilibrium universality class. Specifically, we find that ``classical'' paramagnetic (random-spin) and ``quantum'' paramagnetic (dimerized and antidimerized) initial states yield drastically different critical initial slip exponents in the bulk. This initial-state dependence also extends to boundary critical dynamics, where different initial states yield different surface critical initial slip exponents even within the same boundary universality class. Remarkably, for the antidimerized initial state, we uncover an anomalous scaling law violating the conventional relation between the bulk and surface critical initial slip exponents.
  Our results establish an initial-state-controlled dynamical classification of quantum criticality that extends beyond the conventional equilibrium universality paradigm. 
\end{abstract}

\maketitle

{\bf Introduction}---The concept of universality class provides a fundamental framework for understanding phase transitions and critical phenomena~\cite{Cardy1996book,Sachdev1999book}. Microscopically disparate systems may undergo phase transitions that belong to the same universality class. Different universality classes are characterized by their unique sets of critical exponents, which govern the scaling properties at the corresponding critical points. Identifying the relevant variables and their critical exponents therefore constitutes the central task for establishing the universality class and, in turn, for understanding phase transitions.

In equilibrium phase transitions, the deviation of a tuning parameter from its critical value serves as a relevant scaling variable. Bulk universality classes are then characterized by the scaling dimension of this relevant deviation and the order-parameter field at the critical point. At a spatial boundary, boundary conditions and broken spatial translational symmetry introduce additional relevant scaling variables, yielding different boundary universality classes~\cite{PTCP8,PTCP10,Diehl1997book,Pleimling2004jpa,Binder1972prb,Binder1984prl,Landau1990prb,Ruge1995prb,Deng2005pre,Deng2006pre,Ritschel1995prl,Pleimling2005prb,Pleimling2004prl,Lin2008pre}.
A single bulk universality class may correspond to several distinct boundary universality classes, resulting in a many-to-one structure. Beyond the well-known ordinary, special, surface, and extraordinary boundary transitions, an extraordinary-log universality class has recently been uncovered in O($N$) systems~\cite{Metlitski2022sp,Toldin2021prl,Hu2021prl,Toldin2022prl,Padayasi2022sp,Sun2022prb,Zhang2022prb,Sun2022prb_log,Sun2025sp,Hu2026prl,Toldin2025prr}. Furthermore, in quantum systems, topologically protected edge modes enrich boundary criticality by giving rise to new universality classes~\cite{Zhang2017prl,Zhu2026prl,Wang2024prb,Liu2024prl,Liu2025arx,Toldin2025arx,Wu2020prb,Scaffidi2017prx,Ding2018prl,Weber2018prb,Verresen2021prx,Zhu2021prb,Yu2022prl}.

Beyond equilibrium, critical dynamics represents a long-standing fundamental frontier in quantum many-body physics.
Traditionally, imaginary-time evolution works as a standard approach to ground-state preparation in both numerical simulations~\cite{Vidal2004prl,Jordan2008prl,Jiang2008prl,Assaad2008book,Li2019review,Sandvik2010review} and quantum devices~\cite{Motta2020natphy,Nishi2021npj,Huo2023quantum,Zeng2021,Gluza2026prl,Chang2026prl,Schwartzman2026prxquantum}.
Near a quantum critical point, the imaginary-time relaxation has been shown to exhibit rich nonequilibrium scaling features~\cite{Yin2014prb,Zhang2014pre}.
In particular, starting from the paramagnetic phase with short-range correlation, the temporal correlation function displays a critical initial slip behavior governed by a new dynamic exponent $\theta$~\cite{Yin2014prb,Zhang2014pre}, analogous to the short-time critical dynamics in classical systems~\cite{Janssen1989zpb,Li1995prl,Zheng1996prl,Albano2011rpp}.
Recently, short-imaginary-time critical dynamics has been experimentally verified on quantum-computing hardware~\cite{Zhang2024prb}, and also been used to probe critical properties of more complex condensed-matter systems, including those plagued by the sign problem~\cite{Yin2014prb,Zhang2014pre,Shu2017prb,Shu2022prl,Yu2026prl,Zhang2024prb,Lin2026prb,Yu2026sa,Shen2026nc,Shu2022prb}. 

Motivated by the existence of multiple boundary universality classes associated with a single bulk universality class, it is natural to ask whether the ``temporal surface'', i.e., the short-time regime near the initial state, can likewise support distinct dynamical universality classes within a single equilibrium universality class, as illustrated in Fig.~\ref{motivation}(a).
Although quantum imaginary-time evolution and classical relaxation exhibit similar critical initial slip behavior~\cite{Yin2014prb,Zhang2014pre,Shu2017prb,Shu2022prl,Zhang2024prb,Yu2026prl,Lin2026prb}, the question addressed here is intrinsic to quantum systems. 
For classical systems, the high-temperature paramagnetic phase is typically represented by the random-spin state, and distinct responses arising from different paramagnetic initial states are generally absent. In contrast, quantum paramagnetic states can possess different intrinsic structures. This highlights the unique character of quantum critical dynamics.

In this Letter, we address this question by investigating the imaginary-time relaxation dynamics of the dimerized Heisenberg model, which exhibits a ground-state quantum phase transition (QPT) between the antiferromagnetic (AFM) phase and the quantum paramagnetic phase within the $(2+1)$D Heisenberg universality class, as illustrated in Fig.~\ref{motivation}(b). We consider three representative initial states: a random-spin state, a dimerized state with dimers residing on strong-interaction bonds, and an antidimerized state with dimers on weak-interaction bonds, with short-range correlations and vanishing magnetization.
Although these initial states are all magnetically disordered, their order-parameter temporal correlation functions exhibit strikingly distinct scaling behaviors, governed by different critical initial slip exponents $\theta$, giving rise to distinct initial-state-dictated dynamical universality classes. This initial-state dependence also persists at the boundary, where different initial states yield different surface critical initial slip exponents $\theta_1$ even within the same boundary universality class. Remarkably, for the antidimerized initial state, the boundary autocorrelation function exhibits anomalous scaling that violates the conventional relation between the bulk and surface critical initial slip exponents. 
Our results establish the initial state as an essential role in classifying dynamical universality classes in quantum critical dynamics beyond the equilibrium paradigm. These findings deepen our understanding of dynamical scaling and broaden the scope of universality arguments in quantum many-body nonequilibrium systems.

{\bf Model}---The Hamiltonian of the $2$D columnar dimerized Heisenberg model is given by~\cite{Chakravarty1988prl,Huse1988prl,Singh1989prb,Millis1993prl,Chubukov1994prb}
\begin{equation}
H=J_1 \sum_{\langle ij \rangle}\mathbf{S}_i\cdot\mathbf{S}_j+J_2 \sum_{\langle ij \rangle'}\mathbf{S}_i\cdot\mathbf{S}_j,
\label{eq:hamiltonian}
\end{equation}
in which $\mathbf{S}_i$ denotes the spin-$1/2$ operator at site $i$, and $J_1$ and $J_2$ are the AFM interactions defined on the bonds $\langle ij \rangle$ and $\langle ij \rangle'$, respectively, as illustrated in Fig.~\ref{motivation}(b).

The QPT in this model has been extensively studied as a representative example of the competition between AFM order and quantum singlet formation~\cite{Chakravarty1988prl,Huse1988prl,Singh1989prb,Millis1993prl,Chubukov1994prb,Troyer1996prb,Troyer1998prl,Matsumoto2001prb,Sandvik2006prb,Giamarchi2008natphys,Sachdev2008natphys,Sandvik2010review,Merchant2014natphy,Mila2015prb,Wenzel2008prl,Meng2015prb,Nvsen2018prl,Siqimiao2018prb,Jiang2018prb,Jiang2020prb,Sandvik1994prl,Wenzel2008prl,Wenzel2009prb}.
This competition is also relevant to the magnetic properties of strongly correlated materials, including cuprate superconductors~\cite{McMorrow2001prl,Sachdev2008natphys,RManousakis1991amp,Vojta2007rmp}.
By tuning the coupling ratio $q\equiv J_2/J_1$, the system undergoes a continuous QPT at $q_c=1.90951(1)$, separating the N\'eel AFM phase for $q<q_c$ from the paramagnetic phase for $q>q_c$~\cite{Sandvik2010review,Nvsen2018prl}.
The order parameter characterizing the transition is $ M\equiv L^{-d}\sum_r (-1)^{r_x+r_y}S_r^{z}$,  where $r_{x,y}$ are the coordinates of lattice site $r$ in the $x,y$ directions, respectively.
The QPT belongs to the $(2+1)$D Heisenberg O($3$) universality class, with $\nu=0.7112(5)$, $\beta=0.3689(3)$~\cite{Campostrini2002prb} and $z=1$~\cite{Chakravarty1988prl,Chubukov1994prb,Sachdev2008natphys}, as confirmed by extensive quantum Monte Carlo (QMC) simulations~\cite{Wenzel2009prb,Nvsen2018prl,Ding2018prl}.

In addition, it has been shown that different boundary conditions can give rise to different boundary universality classes~\cite{PTCP8,PTCP10,Diehl1997book,Pleimling2004jpa,Binder1972prb,Binder1984prl,Landau1990prb,Ruge1995prb,Deng2005pre,Deng2006pre,Ritschel1995prl,Pleimling2005prb,Pleimling2004prl,Lin2008pre}.
For boundaries with nondangling spins, the boundary transition belongs to the ordinary universality class with the boundary order parameter exponent $\beta_1=0.825(12)$, while for boundaries with dangling spins, the boundary transition belongs to a nonordinary universality class with $\beta_1=0.1893(9)$, which is close to the exponent of the putative special transition of the Heisenberg model~\cite{Ding2018prl,Weber2018prb,Zhu2021prb}.

We investigate the imaginary-time relaxation dynamics using the projector QMC method~\cite{Sandvik2010review,Grandi2011prb,Grandi2013jpcm,Liu2013prb}. Starting from an initial state $|\psi(0)\rangle$, the system evolves as $|\psi(\tau)\rangle=e^{-\tau H}|\psi(0)\rangle$. The imaginary-time evolution operator is expanded into a power series, and the resulting operator sequences are sampled using local bond and global loop updates. Physical observables are then measured at different imaginary times with the initial state fixed.
Different initial states are implemented through the configurations at the imaginary-time boundary $\tau=0$. To study boundary effects, we consider a $2L\times L$ lattice with open boundary conditions along the $x$ direction and periodic boundary conditions along the $y$ direction.

\begin{figure}[tbp]
\centering
  \includegraphics[width=\linewidth,clip]{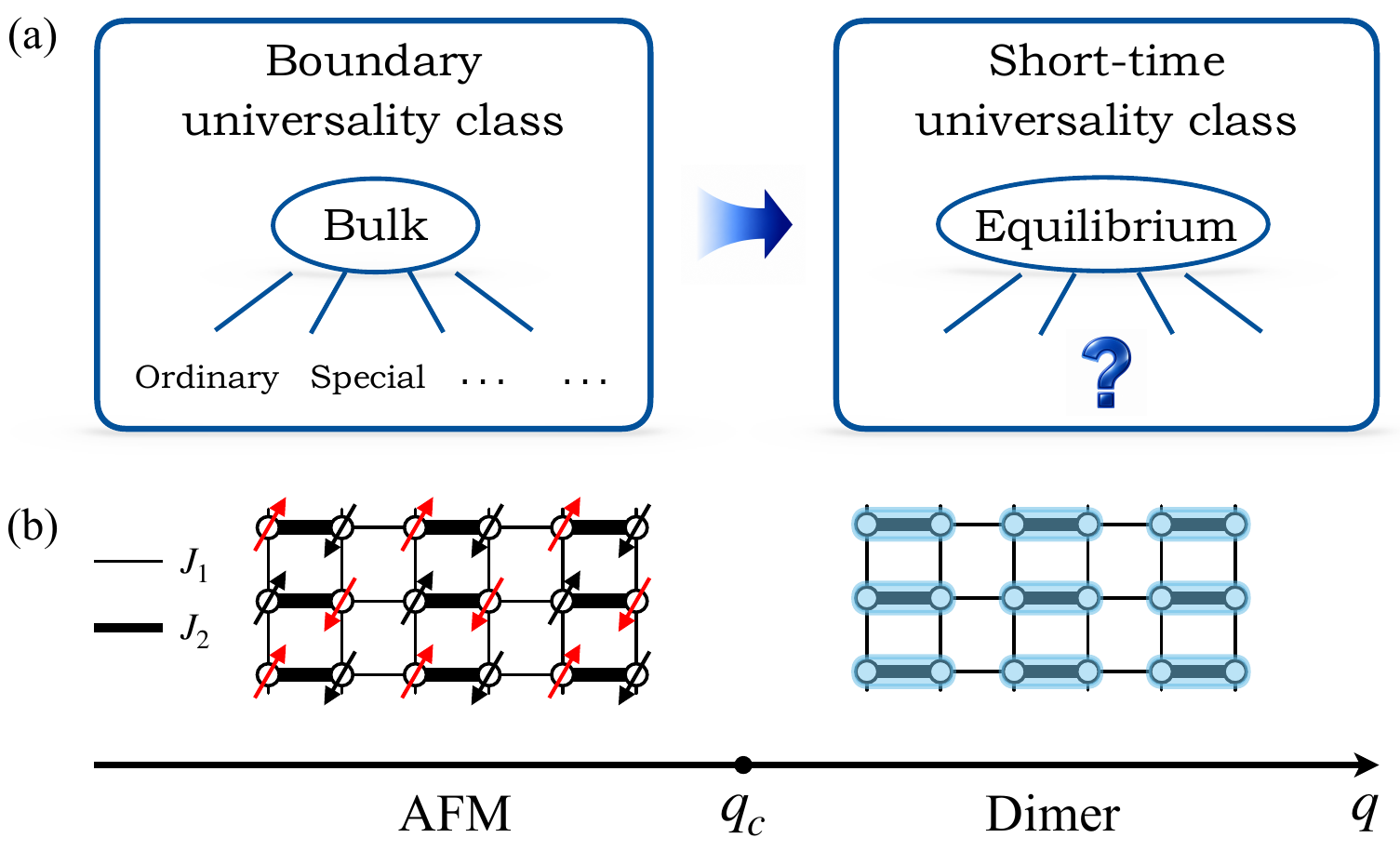}
  \vskip-3mm
  \caption{
    (a) Motivation for classifying short-time dynamical universality classes: whether different initial states can induce different dynamical universality classes within the same equilibrium universality class, analogous to different boundary universality classes associated with the same bulk universality class. 
    (b) Phase diagram of the columnar dimerized Heisenberg model.
  }
  \label{motivation}
\end{figure}

{\bf Short-time critical initial slip}---Near the quantum critical point, the critical initial slip can be characterized by the bulk temporal correlation function $G_b(\tau)\equiv L^{d}\langle M(0)M(\tau)\rangle$~\cite{Shu2017prb}.
For a short-range-correlated initial state, $G_b(\tau)$ obeys the scaling form
\begin{equation}
G_b(\tau,L)=\tau^{\theta}f_1(\tau L^{-z}),
\label{gb}
\end{equation}
where the critical initial slip exponent $\theta$ is an independent dynamic exponent that cannot be determined from equilibrium critical exponents~\cite{Yin2014prb,Zhang2014pre}.

At the boundary, the critical initial slip can instead be characterized by the surface autocorrelation function $A_s(\tau)\equiv\langle S_r^z(0)S_r^z(\tau)\rangle$, with $r$ located at the boundary. Conventionally, $A_s$ obeys the scaling form~\cite{Shu2026arx}
\begin{equation}
A_s(\tau,L)=\tau^{\theta_1-d/z}f_2(\tau L^{-z}),
\label{eq:autoc}
\end{equation}
where $\theta_1$ is the surface initial slip exponent. The short-distance expansion (SDE)~\cite{Symanzik1981npb,PTCP8,PTCP10,Diehl1997book,Ritschel1995prl} relates $\theta_1$ to $\theta$ through
\begin{equation}
\theta_1=\theta-\frac{2(\beta_1-\beta)}{\nu z}.
\label{eq:scarela}
\end{equation}
This scaling law has recently been verified for different boundary universality classes of the quantum Ising model~\cite{Shu2026arx}.

In classical phase transitions, the paramagnetic phase typically corresponds to a high-temperature phase with randomly oriented spins. Accordingly, for a given dynamical equation, $\theta$ bears a one-to-one correspondence with the equilibrium universality class. In quantum systems, however, paramagnetic states can possess different intrinsic structures even when their correlations are short-ranged. 
This raises an important question: Do different paramagnetic initial states fall within the same dynamical universality class, giving identical $\theta$ ($\theta_1$), or may they belong to different dynamical universality classes, leading to different $\theta$ ($\theta_1$)?

As demonstrated below, the latter is indeed the case. Furthermore, we find that the conventional scaling law in Eq.~(\ref{eq:scarela}) can break down for certain initial states.

{\bf Random-spin initial state}--- We study imaginary-time critical relaxation starting from a completely random-spin state with vanishing correlations. This state serves as a representative of a broader class of short-range-correlated ``classical'' paramagnetic initial states exhibiting universal short-time critical behavior.

The scaling behavior of the bulk temporal correlation is shown in Fig.~\ref{random}(a). As seen in the inset, in the short-time regime, $G_b$ shows negligible finite-size dependence and follows $G_b\propto \tau^{-0.45}$, yielding $\theta\simeq -0.45(4)$.
By rescaling $G_b$ and $\tau$ as $G_b L^{-\theta z}$ and $\tau L^{-z}$, respectively, the rescaled curves collapse well, consistent with $G_b(\tau,L)=L^{\theta z}f_3(\tau L^{-z})$. This scaling form is equivalent to Eq.~(\ref{gb}) with $f_3(x)=x^{-\theta}f_1(x)$. These results confirm Eq.~(\ref{gb}) and the value of $\theta$ for the random-spin initial state.

Remarkably, the negative value of $\theta$ contrasts with the positive $\theta$ found in the classical Heisenberg model. The sign of $\theta$ reflects the competition between the growth of AFM correlations and critical fluctuations. While AFM interactions promote the formation and growth of correlated domains, critical fluctuations tend to suppress the development of order. The negative $\theta$ observed here suggests that critical fluctuations dominate the early-time evolution.

Turning to the boundary behavior, the surface autocorrelation function $A_s(\tau)$ exhibits distinct scaling behaviors for different boundary universality classes.
For the ordinary transition without dangling edge spins, the scaling behavior described by Eq.~(\ref{eq:autoc}) is verified in Fig.~\ref{random}(b), yielding $\theta_1\simeq -1.73(5)$, consistent with the scaling relation in Eq.~(\ref{eq:scarela}).
In contrast, for the nonordinary transition associated with dangling edge spins, the analogous analysis in Fig.~\ref{random}(c) yields $\theta_1\simeq 0.055(40)$, again consistent with Eq.~(\ref{eq:scarela}).

\begin{figure}[tbp]
\centering
  \includegraphics[width=\linewidth,clip]{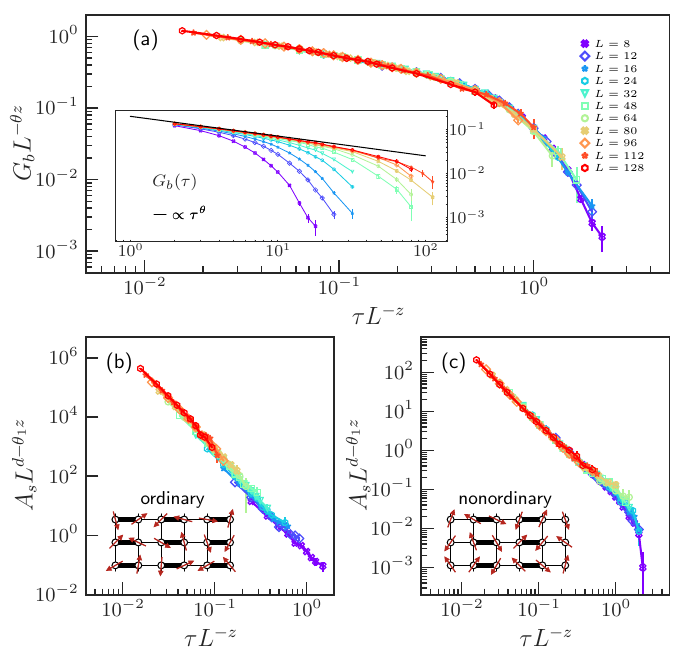}
  \vskip-3mm
  \caption{{\bf Random-spin initial state.}
    (a) Finite-size scaling of the bulk temporal correlation $G_b(\tau)$ with $\theta\simeq 0.45(4)$;the inset shows the unscaled data and the power-law behavior $G_b\propto\tau^\theta$.
    (b) and (c) Finite-size scaling of the surface autocorrelation function $A_s(\tau)$ for the ordinary and nonordinary boundary transitions, respectively, with $\theta_1\simeq -1.73(5)$ and $0.055(40)$. The insets illustrate the corresponding boundary geometries.
  }
  \label{random}
\end{figure}

{\bf Dimerized initial state}--- To unveil whether distinct dynamical universality classes can emerge within the short-time regime, we consider a dimerized initial state consisting of independent dimers residing on the strong bonds. This state represents a class of ``quantum" paramagnetic phases in which weak correlations exist between strong-bond dimers, corresponding to the ground states of the Hamiltonian for $q>q_c$. A key distinction from the random-spin states considered above is that the dimerized state lies in the zero-total-spin sector. Since the Hamiltonian conserves total spin, the imaginary-time evolution remains confined to this sector.

As shown in the inset of Fig.~\ref{dimer}(a), the bulk temporal correlation exhibits a power-law increase, $G_b\propto\tau^{0.28}$, in the short-time regime, yielding $\theta\simeq 0.28(1)$. The data collapse in the main panel further supports this value of $\theta$.
In contrast to the random-spin initial state, for which $\theta<0$, the positive $\theta$ found here signals that domain growth dominates over critical fluctuations.
Physically, total-spin conservation restricts the accessible critical excitations, making domain growth the dominant mechanism and thereby contributing to the positive value of $\theta$. 
The distinct values of $\theta$ demonstrate that different initial paramagnetic states can yield different dynamical universality classes.

The initial-state dependence of dynamical universality classes also persists at the boundary. For the ordinary transition, the scaling collapse of $A_s(\tau)$ yields $\theta_1\simeq -1.00(4)$, as shown in Fig.~\ref{dimer}(b). For the nonordinary transition, the corresponding analysis in Fig.~\ref{dimer}(c) gives $\theta_1\simeq 0.78(1)$. In both cases, the value of $\theta_1$ satisfies the scaling relation in Eq.~(\ref{eq:scarela}), but differs markedly from the corresponding $\theta_1$ for the random-spin initial state. Thus, different initial states give rise to distinct dynamical universality classes even within the same boundary universality class.

\begin{figure}[tbp]
\centering
  \includegraphics[width=\linewidth,clip]{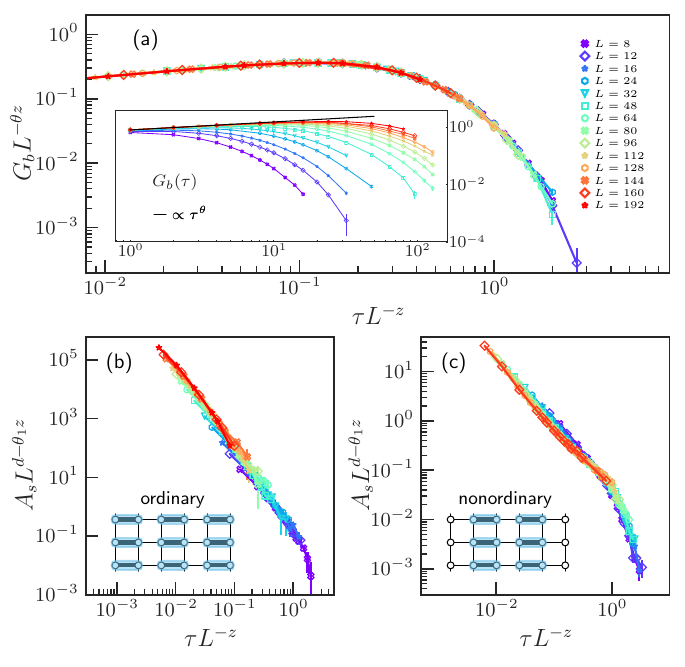}
  \vskip-3mm
  \caption{{\bf Dimerized initial state.}
    (a) The bulk temporal correlation exhibits critical initial slip with $\theta\simeq 0.28(1)$.
    (b) and (c) Scaling of the surface autocorrelation $A_s(\tau)$ at the ordinary and nonordinary transitions, yielding $\theta_1\simeq -1.00(4)$ and $0.78(1)$, respectively. The insets illustrate the corresponding initial-state configurations.
  }
  \label{dimer}
\end{figure}

{\bf Antidimerized initial state}---We further consider an antidimerized initial state, in which singlet dimers occupy the weak bonds, opposite to the strong-bond dimerization characteristic of the quantum paramagnetic phase.

Since this antidimerized state differs from the dimerized state only in the spatial rearrangement of dimers, one may reasonably expect similar critical dynamics. However, the observed behavior is strikingly different: as shown in Fig.~\ref{antidimer}(a1), the bulk correlation $G_b$ first exhibits a burst-like exponential growth, reaches a peak, and then decays. This stands in sharp contrast to the power-law evolution observed for the two initial states considered earlier. 

\begin{figure}[tbp]
\centering
  \includegraphics[width=\linewidth,clip]{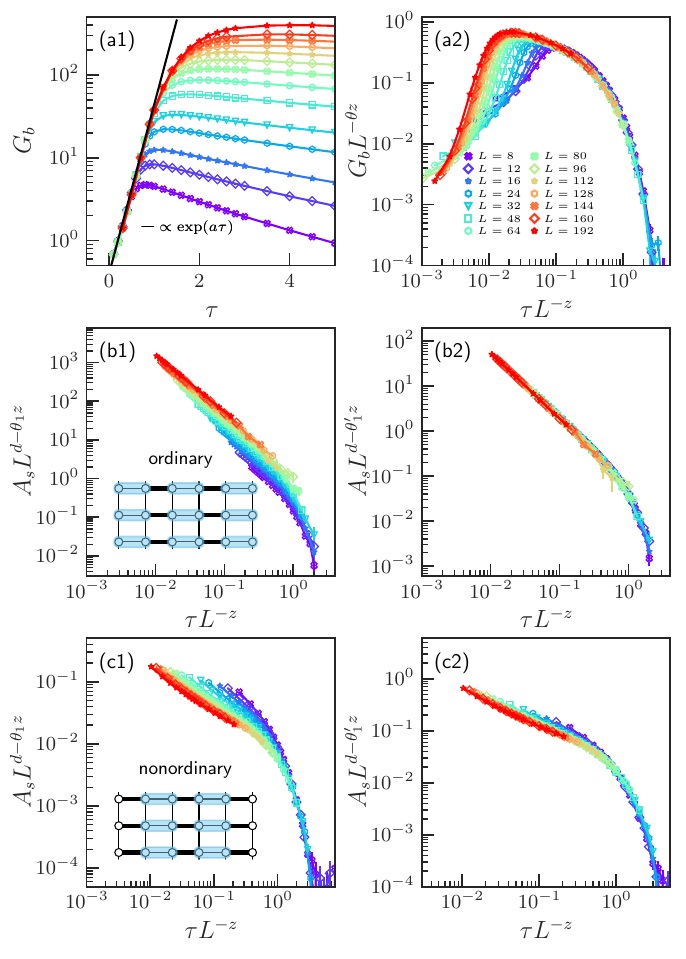}
  \vskip-3mm
  \caption{{\bf Antidimerized initial state}
    (a1) Burst-like exponential growth of $G_b(\tau)$ at early times.
    (a2) Critical scaling of $G_b(\tau)$ with $\theta\simeq 1.21(1)$.
    (b1) and (b2) Scaling collapse of $A_s$ for the ordinary transition: No collapse observed for the conventional scaling with $\theta_1\simeq -0.07(4)$, and successful collapse with the anomalous exponent $\theta_1'\simeq 0.57(2)$, respectively.
    (c1) and (c2) Corresponding scaling collapse for the nonordinary transition with $\theta_1\simeq1.72(1)$ and $\theta_1'\simeq 1.46(1)$, repspectively. The insets depict the initial-state configurations.
  }
  \label{antidimer}
\end{figure}

The rapid early growth can be attributed to the mismatch between the antidimerized initial state and the interactions in the Hamiltonian. The imaginary-time evolution with the Hamiltonian~(\ref{eq:hamiltonian}) tends to build up correlations between spins linked by the strong bonds, while these very spins already form entangled singlet pairs with the spins at the other ends of the weak bonds in the initial state. As a result, the whole system becomes correlated during the early stage of evolution.
It is worth noting that this process is dominated by the interplay between microscopic interactions and the initial-state structure, rather than critical fluctuations.
Thus, the early-time behavior of $G_b$ lies outside the critical scaling regime. Nevertheless, the correlations developed during this stage can influence the subsequent critical dynamics.

After the initial exponential growth, $G_b$ crosses over into the critical scaling regime. As shown in Fig.~\ref{antidimer}(a2), the scaling behavior of $G_b$ is well described by Eq.~(\ref{gb}) with $\theta\simeq 1.21(1)$.
This value is markedly different from those obtained for the other two initial states, indicating a distinct dynamical universality class.

We next consider the boundary dynamics for the antidimerized initial state. For the preceding two initial states, the boundary exponent $\theta_1$ satisfies the bulk-boundary relation in Eq.~(\ref{eq:scarela}).
However, the rescaled $A_s(\tau)$ fails to collapse for both the ordinary and nonordinary transitions when $\theta_1$ is determined from this relation, as shown in Figs.~\ref{antidimer}(b1) and (c1). Instead, a good scaling collapse is obtained by rescaling $A_s$ and $\tau$ as $A_sL^{d-\theta_1'z}$ and $\tau L^{-z}$, respectively, with $\theta_1'\simeq 0.57(2)$ for the ordinary transition and $\theta_1'\simeq 1.46(1)$ for the nonordinary transition, as shown in Figs.~\ref{antidimer}(b2) and (c2). 
Remarkably, these exponents are consistent with a modified bulk–boundary scaling law,
\begin{equation}
\theta_1'=\theta-\frac{\beta_1-\beta}{\nu z}.
\label{eq:scarela1}
\end{equation}
This reveals a breakdown of the conventional bulk-boundary scaling law of Eq.~(\ref{eq:scarela}) for the antidimerized initial state.

To understand this anomalous scaling law, we revisit the relations among $\theta_1$, $\theta_1'$, and $\theta$ from the perspective of SDE.
The surface autocorrelation function $A_s(\tau)$ involves two boundary order-parameter operators $S_r^z(0)$ and $S_r^z(\tau)$. In the conventional case, the SDE applies separately to the two operators, giving rise to the factor of $2$ in Eq.~(\ref{eq:scarela}). For the antidimerized initial state, however, the initial burst rapidly builds up correlations throughout the system and modifies the role of the operator at $\tau=0$. Consequently, the SDE applies only on the operator $S_r^z(\tau)$, reducing the boundary contribution from $2(\beta_1-\beta)/\nu z$ to $(\beta_1-\beta)/\nu z$. This provides a plausible explanation for the scaling relation in Eq.~(\ref{eq:scarela1}). This interpretation will be further corroborated by additional physical quantities presented in the Supplementary Materials.

{\bf Discussion}---The results above demonstrate that distinct initial states can give rise to different short-time dynamical universality classes, characterized by different critical initial slip exponents.
This initial-state dependence is a distinctive feature of quantum critical dynamics, rooted in the rich internal structure of quantum disordered states, including their conserved quantities and preformed correlations.
Near the quantum critical point, critical slowing down allows the influence of these structures to persist over long imaginary-time scales, thereby leading to different dynamical universality classes. In this respect, the initial state plays a role analogous to boundary conditions in equilibrium critical phenomena, where the interplay between bulk criticality and different boundary conditions can give rise to different boundary universality classes.

However, the classification of initial-state-dictated dynamical universality classes is fundamentally different from that of spatial boundary criticality. Although the initial state can be regarded as a boundary along the time direction, time evolution obeys intrinsic causality: the initial state affects subsequent dynamics, whereas late-time evolution cannot alter the initial state. This contrasts with spatial boundaries, which interact mutually with the bulk. Moreover, initial states carrying nontrivial conserved quantities often do not have direct counterparts of boundary states under the quantum–classical correspondence. Therefore, the initial state furnishes an independent classification scheme for dynamical universality classes, opening up a distinct avenue for exploring nonequilibrium critical phenomena in quantum many-body systems.

In the Supplemental Materials, we consider additional configurations with different boundary conditions and show that their critical dynamics are consistent with the classification established above. Although the different critical initial slip exponents provide evidence for initial-state-dictated dynamical universality classes, their classification within a field-theoretical framework remains an open question. Furthermore, this work provides important ingredients for studying imaginary-time quantum criticality in rapidly developing quantum devices and holds promise for its realization therein~\cite{Zhang2024prb}.

\begin{table}[h]
\caption{Different critical initial slip exponents for different initial-state-dictated dynamical universality classes.}
\label{tablecom}
\begin{ruledtabular}
\begin{tabular}{cccc}
 \multirow{2}{*}{Initial state} & \multirow{2}{*}{Bulk} & \multicolumn{2}{c}{Boundary} \\
\cline{3-4}
& & Ordinary & Nonordinary \\
\hline
Random-spin & $-0.45(4)$ & $-1.73(5)$ & $0.055(40)$ \\
Dimerized & $0.28(1)$ & $-1.00(4)$ & $0.78(1)$ \\
Antidimerized & $1.21(1)$ & $0.57(2)$ & $1.46(1)$ \\
\end{tabular}
\end{ruledtabular}
\end{table}

{\bf Summary}---We have studied the short-imaginary-time critical dynamics of the dimerized Heisenberg model, which undergoes a QPT between the AFM and quantum paramagnetic phases in the $(2+1)$-dimensional Heisenberg universality class. Starting from three representative disordered initial states---a random-spin state, a dimerized state, and an antidimerized state---we demonstrate that different types of initial states can induce distinct dynamical universality classes, characterized by different critical initial slip exponents $\theta$ of the bulk temporal correlation function. 
This initial-state dependence also persists at the boundary, where different initial states yield different surface critical initial slip exponents even within the same boundary universality class, as summarized in Table~I. 
In particular, for the antidimerized initial state, the conventional bulk–boundary scaling law breaks down, with the surface autocorrelation instead governed by $\theta_1'=\theta-(\beta_1-\beta)/\nu z$.
This anomalous scaling law may be associated with the burst-like growth of correlations at early imaginary times, which modifies the role of the initial boundary operator in the subsequent critical dynamics.
Our results establish the initial state as an independent ingredient in determining dynamical universality classes. Such initial-state dependence is intrinsic to quantum critical dynamics and has no counterpart in conventional classical critical dynamics, where the paramagnetic state is unique, and goes beyond the conventional quantum–classical correspondence.

{\bf Acknowledgments}---We thank Zi-Xiang Li and Yin-Kai Yu for helpful discussions. S.Y. is supported by the National Natural Science Foundation of China (Grants No. 12222515 and No. 12075324), the Research Center for Magnetoelectric Physics of Guangdong Province (Grant No. 2024B0303390001), the Guangdong Provincial Key Laboratory of Magnetoelectric Physics and Devices (Grant No. 2026B1212170049), and Quantum Science and
Technology-National Science and Technology Major Project (Grant No.2025ZD0300400). 
Y.R.S. is supported by the National Natural Science Foundation of China (Grant No. 12104109), the Science and Technology Projects in Guangzhou (Grant No. 2024A04J2092).

\textcolor{black}{\it Data availability}---The data that support the findings of this article are openly available.

\bibliographystyle{apsrev4-2}
\let\oldaddcontentsline\addcontentsline
\renewcommand{\addcontentsline}[3]{}
\bibliography{ref.bib}
\let\addcontentsline\oldaddcontentsline
\onecolumngrid

\clearpage
\newpage
\widetext

\begin{center}
\textbf{\large Supplemental Materials for ``Initial-State-Dictated Dynamical Universality Classes in Quantum Criticality''}
\end{center}

\date{\today}
\maketitle

\renewcommand{\thefigure}{S\arabic{figure}}
\setcounter{figure}{0}
\renewcommand{\theequation}{S\arabic{equation}}
\setcounter{equation}{0}
\renewcommand{\thesection}{\Roman{section}}
\setcounter{section}{0}
\setcounter{secnumdepth}{4}

\makeatletter
\renewcommand{\numberline}[1]{\hb@xt@3em{#1\hfil}}
\makeatother
\addtocontents{toc}{\protect\setcounter{tocdepth}{0}}
{
\tableofcontents
}
\section{Initial-State and Boundary Configurations}

\begin{figure}[tbp]
\centering
  \includegraphics[width=0.85\linewidth,clip]{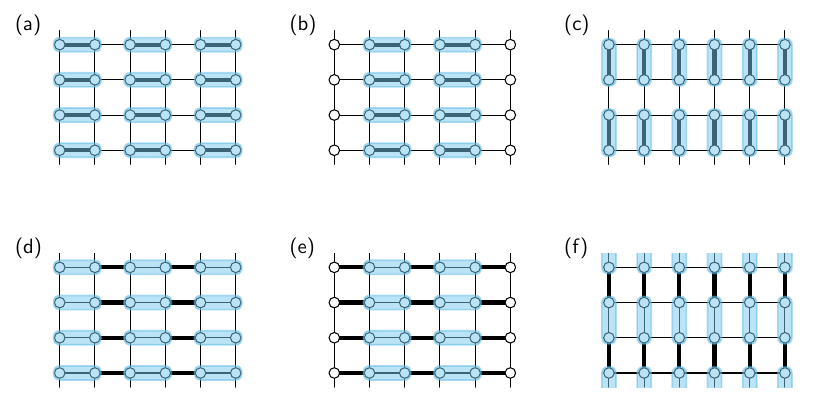}
  \vskip-3mm
  \caption{Six initial-state and boundary configurations considered in
this work. Panels (a)--(c) show dimerized initial states, and panels
(d)--(f) show the corresponding antidimerized initial states.
The thick bonds denote the strong couplings in the Hamiltonian, and the shaded ellipses indicate singlet pairs in the initial state.
The boundaries are open along $x$ and periodic along $y$.
Configurations (c) and (f) have strong bonds and dimers along the $y$ direction, while in other configurations, they are along the $x$ direction.
}
  \label{figs:configs}
\end{figure}

In the main text, we consider three different types of initial states, including the random-spin, dimerized, and antidimerized states. For the latter two states, additional configurations can be constructed by changing the arrangement of the initial singlets and the orientation of the strong bonds. Here we consider six configurations, as illustrated in Fig.~\ref{figs:configs}. Configurations (a), (b), (d), and (e) have already been introduced in the main text and are included here for completeness, as we will present additional results for these configurations in the following sections.

The system has a size of $2L\times L$, with open boundary conditions along the $x$ direction and periodic boundary conditions along the $y$ direction. The two open boundaries are located at $x=1$ and $x=2L$. The upper row of Fig.~\ref{figs:configs} shows three configurations of the dimerized initial state, while the lower row shows the antidimerized configurations.
Configurations (a) and (d) belong to the ordinary boundary universality class, as there are no dangling spins at the open boundaries. In contrast, configurations (b) and (e) have dangling spin chains at $x=1$ and $x=2L$, falling in to nonordinary boundary universality class.
In configurations (c) and (f), the strong bonds are oriented along the $y$ direction, forming vertical dimers at the open boundaries. Consequently, these configurations also belong to the ordinary boundary universality class. The initial singlet arrangement in configuration (f) is shifted by one lattice spacing along the periodic $y$ direction relative to that in configuration (c).

\section{Bulk Critical Dynamics}

We first compare the bulk order-parameter temporal correlation function $G_b(\tau,L)$
for all six configurations.
As a global observable, $G_b$ is governed by bulk critical dynamics, with boundary effects becoming negligible in the thermodynamic limit. We therefore expect the different configurations with the same type of initial state to exhibit identical universal scaling behavior.

\begin{figure}[tbp]
\centering
  \includegraphics[width=0.85\linewidth,clip]{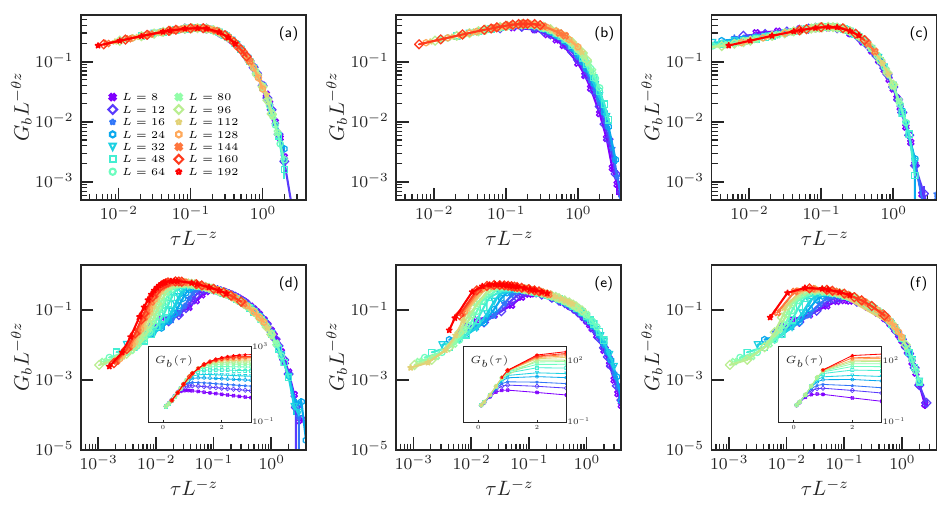}
  \vskip-3mm
  \caption{Finite-size scaling of the bulk correlation function
$G_b$ for the six configurations in Fig.~\ref{figs:configs}.
Panels (a)-(f) correspond to configurations (a)-(f), respectively.
The data are plotted as $G_bL^{-\theta z}$ versus $\tau L^{-z}$,
using $z=1$, $\theta\simeq0.28(1)$ for the dimerized initial states
(a)-(c), and $\theta\simeq1.21(1)$ for the antidimerized initial states
(d)-(f). The insets in (d)-(f) show the unscaled $G_b(\tau)$, exhibiting a burst-like exponential growth at short imaginary times,
}
  \label{figs:smgb}
\end{figure}

As shown in Fig.~\ref{figs:smgb}, the bulk correlation $G_b$ exhibits good finite-size scaling collapse for all six configurations after rescaling $G_b$ and $\tau$ as $G_bL^{-\theta z}$ and $\tau L^{-z}$, respectively. 
As shown in Figs.~\ref{figs:smgb}(a)–(c), the data for the dimerized initial states exhibit good scaling collapse with $\theta\simeq0.28(1)$. For the antidimerized initial states, Figs.~\ref{figs:smgb}(d)–(f) show that the data also collapse well with $\theta\simeq1.21(1)$, even though $G_b$ exhibits a burst-like exponential growth at short imaginary times, followed by a peak and subsequent decay, in sharp contrast to the power-law evolution observed for the dimerized initial states.
Thus, the two initial-state classes exhibit distinct bulk dynamics across the boundary geometries considered here.

\section{Surface Critical Dynamics for the Dimerized Initial State}

To further investigate the surface critical dynamics, we consider three two-time correlation functions~\cite{Shu2026arx}. We first revisit the surface autocorrelation function $A_s$, which has been studied in the main text and is defined as
\[
A_s(\tau)=\langle S_r^z(\tau)S_r^z(0)\rangle,
\]
where $r$ is located at the boundary. Its finite-size scaling form is given by
\[
A_s(\tau,L)=L^{\theta_1 z-d}f_A(\tau L^{-z}),
\]
where $\theta_1=\theta-2(\beta_1-\beta)/\nu z$ is the boundary initial slip exponent.

Beyond the single-site autocorrelation, we consider the local order-parameter correlation function
\[
P_s(\tau)=L^{d-1}\langle M_s(\tau)M_s(0)\rangle,
\]
where $M_s$ denotes the local order parameter obtained by averaging spins along the boundary. Since $P_s$ involves correlations over the entire boundary line, its scaling form differs from that of $A_s$ by a factor associated with the spatial integration,
\[
P_s(\tau,L)=L^{\theta_1 z-1}f_P(\tau L^{-z}).
\]
Both $A_s$ and $P_s$ involve two boundary operators and are therefore governed by the same exponent $\theta_1$.

We further examine the mixed two-time correlation function
\[
G_s(\tau)=L^{d-1}\langle M_s(\tau)M(0)\rangle,
\]
where $M(0)$ is the initial global order parameter averaged over the entire lattice. Unlike $P_s$, which correlates the boundary order parameter at two different imaginary times, $G_s$ measures its correlation with the initial global order parameter. Replacing $M_s(\tau)$ with the global order parameter $M(\tau)$ recovers the bulk correlation function $G_b$ discussed in the main text. Since only one of the two operators is located at the boundary, the SDE contributes only once, leading to
\[
G_s(\tau,L)=L^{\theta_1' z}f_G(\tau L^{-z}),
\]
where $\theta_1'=\theta-(\beta_1-\beta)/\nu z$.

These scaling forms indicate that $A_s$ and $P_s$ are governed by $\theta_1$, whereas $G_s$ is governed by $\theta_1'$, providing complementary probes of the boundary critical initial slip.

\begin{figure}[tbp]
\centering
  \includegraphics[width=0.85\linewidth,clip]{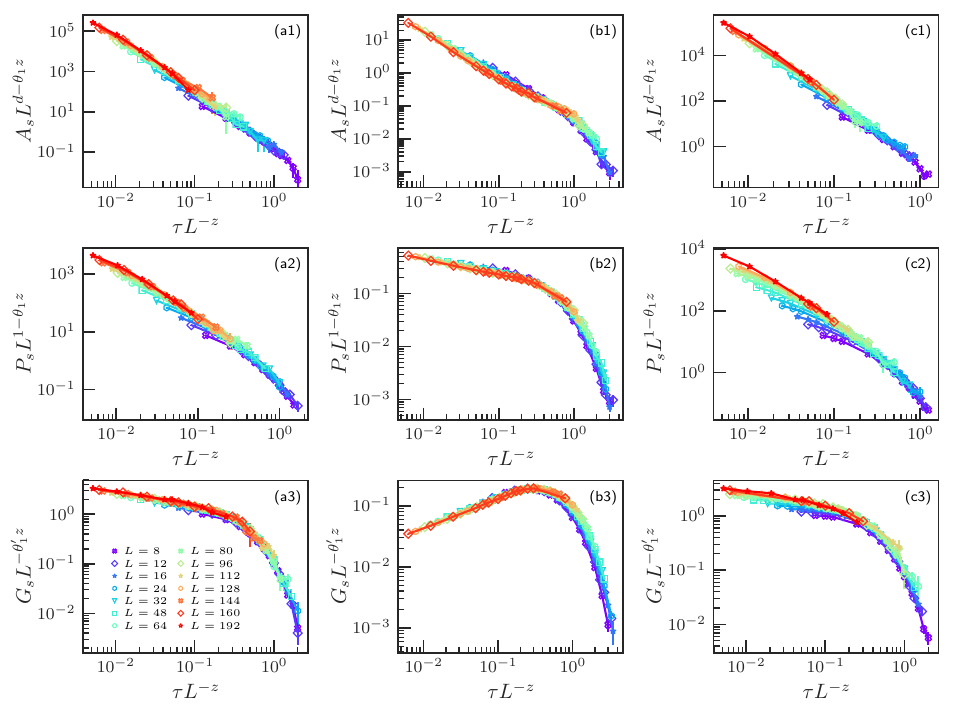}
  \vskip-3mm
  \caption{Finite-size scaling of surface correlations for the dimerized initial states. The three columns correspond to configurations (a)-(c) in Fig.~\ref{figs:configs}, with ordinary boundaries in the first and third columns and nonordinary boundaries in the second. From top to bottom, the panels show the rescaled $A_sL^{d-\theta_1z}$, $P_sL^{1-\theta_1z}$, and $G_sL^{-\theta_1'z}$, respectively, as functions of $\tau L^{-z}$. The scaling collapse is obtained using $\theta_1\simeq-1.00(4)$ and $\theta_1'\simeq-0.36(2)$ for the ordinary class, and $\theta_1\simeq0.79(1)$ and $\theta_1'\simeq0.53(1)$ for the nonordinary class.
}
  \label{figs:smdimer}
\end{figure}

Figure~\ref{figs:smdimer} presents the finite-size scaling results for $A_s$, $P_s$, and $G_s$ for the dimerized initial states. The three columns correspond to configurations (a), (b), and (c) in Fig.~\ref{figs:configs}, respectively, while the three rows show $A_s$, $P_s$, and $G_s$. The first and third columns belong to the ordinary boundary universality class, with $\theta_1\simeq-1.00(4)$ and $\theta_1'\simeq-0.36(2)$, whereas the second column corresponds to the nonordinary class, with $\theta_1\simeq0.79(1)$ and $\theta_1'\simeq0.53(1)$. As shown in the first two rows, the data for $A_s$ and $P_s$ exhibit good finite-size scaling collapse using $\theta_1$, while the results for $G_s$ in the third row are consistent with the scaling governed by $\theta_1'$. These results support the conventional boundary initial slip scaling relations for the dimerized initial states across different boundary geometries.

\section{Surface Critical Dynamics for the Antidimerized Initial State}

\begin{figure}[tbp]
\centering
  \includegraphics[width=0.85\linewidth,clip]{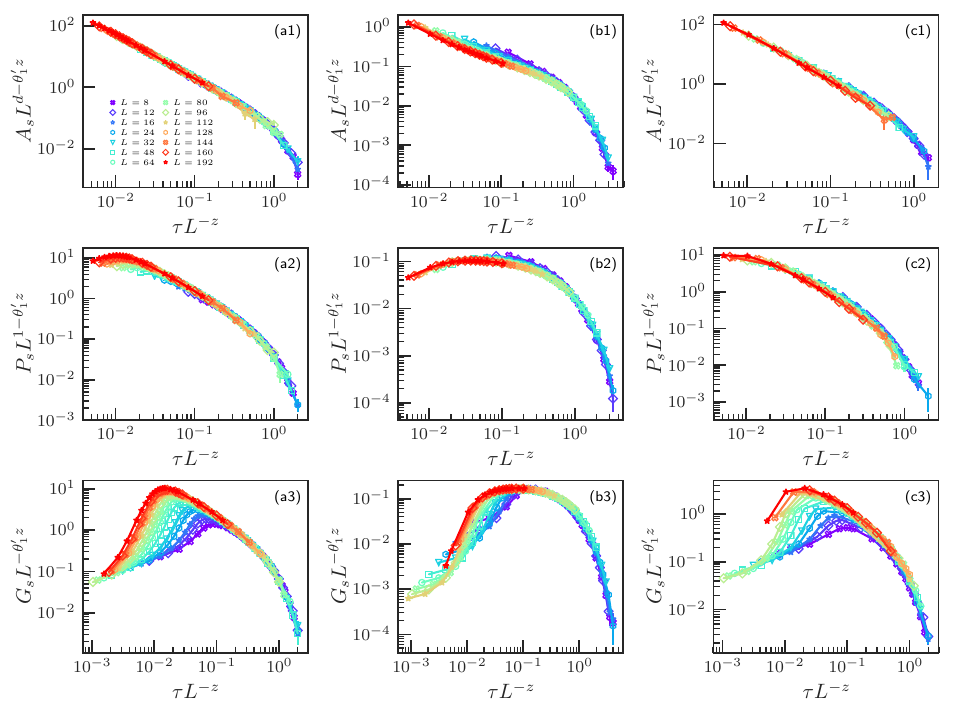}
  \vskip-3mm
  \caption{Finite-size scaling of surface correlations for the antidimerized initial states. The three columns correspond to configurations (d)-(f) in Fig.~\ref{figs:configs}, respectively, with ordinary boundaries in the first and third columns and nonordinary boundaries in the second. From top to bottom, the panels show the rescaled $A_sL^{d-\theta_1'z}$, $P_sL^{1-\theta_1'z}$, and $G_sL^{-\theta_1'z}$, respectively, as functions of $\tau L^{-z}$. All three quantities exhibit good scaling collapse using $\theta_1'\simeq0.57(2)$ for the ordinary class and $\theta_1'\simeq1.46(1)$ for the nonordinary class.}
  \label{figs:smadimer}
\end{figure}

We now turn to the surface critical dynamics for the antidimerized initial states.
Figure~\ref{figs:smadimer} presents the finite-size scaling results for $A_s$, $P_s$, and $G_s$ in configurations (d)-(f). The three columns correspond to configurations (d), (e), and (f), respectively, with ordinary boundary criticality in the first and third columns and nonordinary boundary criticality in the second.

Remarkably, all three correlation functions exhibit good scaling collapse using the same boundary exponent $\theta_1'=\theta-(\beta_1-\beta)/\nu z$, with $\theta_1'\simeq0.57(2)$ for the ordinary class and $\theta_1'\simeq1.46(1)$ for the nonordinary class. In particular, the scaling of $A_s$ and $P_s$ is governed by $\theta_1'$ rather than the conventionally expected exponent $\theta_1=\theta-2(\beta_1-\beta)/\nu z$ as in the other two types of initial states. This behavior is observed for both the ordinary and nonordinary boundary universality classes, and persists when the strong bonds are oriented along the $y$ direction.

The short-time evolution of these surface correlations exhibits distinct behaviors. While $A_s$ and $P_s$ follow power-law behavior, $G_s$ shows a burst-like growth followed by a decay, similar to the bulk correlation $G_b$. Despite this difference, all three quantities exhibit good finite-size scaling collapse governed by $\theta_1'$ rather than the conventional scaling relation for $A_s$ and $P_s$. The consistent results across the three configurations further support the universality of the anomalous boundary critical dynamics associated with the antidimerized initial state.

In summary, the additional configurations confirm that the distinct dynamical scaling behaviors of the dimerized and antidimerized initial states are robust against changes in boundary geometry and strong-bond orientation. Although the boundary critical exponents differ between the ordinary and nonordinary transitions, the dimerized initial states consistently obey the conventional boundary initial slip scaling relations, whereas the antidimerized initial states exhibit the anomalous scaling relation. These results further support the conclusion of the main text that the internal structure of quantum disordered initial states gives rise to distinct dynamical universality classes within the same equilibrium universality class.

\end{document}